\documentclass[english]{article}
\usepackage[T1]{fontenc}
\usepackage[latin1]{inputenc}
\usepackage{geometry}
\usepackage{graphicx}

\makeatletter

\providecommand{\tabularnewline}{\\}

\usepackage{graphicx}

\AtBeginDocument{

}

\usepackage{babel}
\makeatother
\begin{document}

\title{\textbf{Relativistic and correlation effects in atoms}}

\author{B. P. Das, K. V. P. Latha, Bijaya K. Sahoo, Chiranjib Sur, Rajat
K Chaudhuri \\
\emph{Atomic and Molecular Physics Group, Indian Institute of Astrophysics,
}\\
\emph{Bangalore - 560 034, India}\\
 and\\
D. Mukherjee\\
\emph{Indian Association for the Cultivation of Science, Kolkata -
700 032, India}}

\maketitle
\begin{abstract}
This review article deals with some case studies of relativistic and
correlation effects in atomic systems. After a brief introduction
to relativistic many-electron theory, a number of applications ranging
from correlation energy to parity non-conservation in atoms are considered.
There is a special emphasis on relativistic coupled-cluster theory
as most of the results presented here are based on it.

\end{abstract}

\section{Introduction}

One of the most important milestones in the development of theoretical
atomic physics has been the formulation and application of theories
that can simultaneously treat relativistic and correlation effects
in atoms. Following the early work of Swirles on relativistic Hartree-Fock
or Dirac-Fock (DF) theory \cite{swirles}, Grant made pioneering contributions
to the numerical and angular momentum aspects of this theory \cite{grant-intro}
which paved the way for further advances in the field. Multi-configuration
Dirac-Fock (MCDF) \cite{desclaux-intro,mcdf-grant} and relativistic
many-body perturbation theory (RMBPT) \cite{andrie-unpubl} codes
were developed in the mid 1970s and early 1980s calculations based
on them soon followed \cite{rmbpt-andriessen,desclaux-cheng,grant-scrip,das-andrie}.
During the 1980s and 1990s, these two theories were applied to a wide
range of atoms and ions to study a variety of properties \cite{verhey,blundell,hartley,saperstein}.
The extension of coupled-cluster theory to the relativistic regime
during the last decade is indeed a very significant development \cite{kaldor-book}.
Linear and non-linear versions of this theory have been successfully
used in performing high precision calculations of a number of different
atomic properties \cite{blundell-johnson,kaldor,solomonson-pra91,geetha-2002}.

The present review is by no means comprehensive; it mainly highlights
some of the work on relativistic and correlation effects in atoms
undertaken in our group. Unlike molecules, a number of different relativistic
many-body calculations have been carried out on atoms using a variety
of methods. Relativistic many-body calculations on atoms are currently
much more advanced than those on molecules \cite{mol-1,mol-2,mol-3,mol-4}.
In addition to the inclusion of the Breit interaction, certain types
of QED effects have also been included in atomic calculations. It
will take several years before molecular calculations reach this level
of sophistication. Relativistic many-body calculations of parity and
time reversal violations in some atoms have been performed to an accuracy
of better than 1\%. These calculations in combination with accurate
experiments are now poised to test the Standard Model (SM) of particle
physics. It is not clear at the present time whether it would be possible
to achieve something comparable from studies on discrete symmetry
violations of molecules.

The organization of the paper is as follows : Section \ref{dirac-coul}
deals with the Dirac-Coulomb approximation and the section following
it (section \ref{beyond-DC}), touches upon the Breit interaction
and QED effects. Section \ref{CC-Theory} is an overview of relativistic
coupled-cluster theory which has been used in the majority of the
calculations considered here. In section \ref{Rel-Eff}, we present
the basic ideas underlying two physical effects that are relativistic
in origin -- fine-structure splitting and permanent electric dipole
moment of atoms arising from the electric dipole moment of an electron
and have given the results of some representative calculations. The
enhancement of relativistic effects in heavy atoms along with the
influence of electron correlation is discussed in section \ref{rel-enhance}
with reference to correlation energy, hyperfine interactions and parity
non-conservation in atoms. In the last section we make some concluding
remarks.

\section{\label{dirac-coul}The Dirac-Coulomb Approximation}

For an $N$-electron atom, the relativistic Hamiltonian is given by

\begin{equation}
H=\sum_{i=1}^{N}\left[c\vec{\alpha_{i}}\cdot\vec{p}_{i}+\beta mc^{2}+V_{N}(r_{i})\right]+\sum_{i<j}^{N}\frac{e^{2}}{r_{ij}}\,,\label{dc}\end{equation}
 where $\alpha$ and $\beta$ are given by $\alpha_{i}=\left(\begin{array}{cc}
0 & \sigma_{i}\\
\sigma_{i} & 0\end{array}\right)$ and $\beta=\left(\begin{array}{cc}
I & 0\\
0 & -I\end{array}\right)$; $\sigma_{i}$ are the Pauli matrices and $I$ represents the unit
matrix. $V_{N}(r_{i})$ is the nuclear potential at the site of the
$i$th electron and the last term is the Coulomb interaction between
the electrons. $H$ defined above is known as Dirac-Coulomb Hamiltonian
which is clearly not covariant.

This Hamiltonian can also be written as 

\begin{equation}
H=\sum_{i}h_{0}(i)+\sum_{i<j}\frac{e^{2}}{r_{ij}}\,.\label{dc-1}\end{equation}

The electron-electron interaction can be approximated by an average
potential where each electron moves independently in an average field
caused by the nucleus and the other electron. This is the independent
particle model which is the starting point of most atomic physics
calculations. This can be put into a mathematical footing by partitioning
the full Hamiltonian in the following way : \begin{equation}
H=H_{0}+V_{es},\label{ipm}\end{equation}
 where

\begin{equation}
H_{0}=\sum_{i}^{N}h_{0}(i)\label{h0}\end{equation}
 is a sum of the one electron operators, \begin{equation}
h_{0}(i)=c\alpha_{i}\cdot p_{i}+\beta mc^{2}+U(r_{i}).\label{dc-sing}\end{equation}
 It is customary to assume $U$ as the Dirac-Fock potential \cite{strange-QM}
and \begin{equation}
V_{es}=-\sum_{i}U(r_{i})+\sum_{i<j}\frac{e^{2}}{r_{ij}}\label{v-res}\end{equation}
 can be treated as a perturbation if there are no strongly interacting
configurations in the system. The many-body atomic state $|\Psi(\Gamma,J,M)\rangle$
is an eigen function of the Dirac-Coulomb Hamiltonian and satisfies
the equation, \begin{equation}
H|\Psi(\Gamma,J,M)\rangle=E|\Psi(\Gamma,J,M)\rangle,\label{many-body}\end{equation}
 where $J,M$ are the total angular momentum quantum numbers and $\Gamma$
is the quantum number which distinguishes each of the atomic states.
These states are expanded in terms of the determinantal wavefunctions
which in turn are built from the single particle orbitals. If $|\Phi(\Gamma,J,M)\rangle$'s
denote the determinantal wavefunctions, then, \begin{equation}
|\Psi(\Gamma,J,M)\rangle=\sum_{k}C_{k}|\Phi_{k}(\Gamma,J,M)\rangle.\label{csf}\end{equation}
The coefficients $C_{k}$s are determined by the choice of the theory.
The single particle orbitals are the two-component Dirac spinors, 

\begin{equation}
|\phi_{n\kappa m}\rangle=\frac{1}{r}\left(\begin{array}{c}
P_{n\kappa}(r)\chi_{\kappa m}\\
iQ_{n\kappa}(r)\chi_{-\kappa m}\end{array}\right),\label{dirac-spinor}\end{equation}
where $n$ and $m$ are the principal quantum number and magnetic
quantum number respectively. $\kappa$ is a quantum number given by

\begin{equation}
\kappa=\left\{ \begin{array}{c}
l\,\,\,\,\mathrm{for}\, j=l-\frac{1}{2}\\
-(l+1)\,\,\,\,\,\mathrm{for}\, j=l+\frac{1}{2}\end{array}\right.,\label{kappa}\end{equation}
where $l$ is the orbital angular momentum and $j$ is the total angular
momentum of an electron.

\begin{figure}
\begin{center}\includegraphics{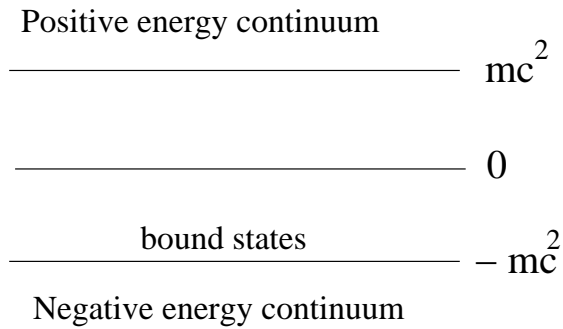}\end{center}

\caption{\label{energy-level}Positive and negative energy states}
\end{figure}

The solutions of the Dirac equation admit both positive and negative
energy states \cite{strange-QM} and this is shown in figure \ref{energy-level}.
For a free particle, only continuum states exist above $mc^{2}$and
below $-mc^{2}$. However, electrons in an atom that are acted on
by a relativistic mean-field potential in addition to continuum states
above $mc^{2}$ and below $-mc^{2}$bound states do exist in the interval
$-mc^{2}$ and $mc^{2}$. The variational principle fails due to the
presence of negative energy states \cite{stanton-jcp84}. The radial
parts of the large and small components of the Dirac spinor are expanded
in terms of Gaussian functions \cite{mohanty-clementi} as follows

\[
P_{n\kappa}(r)=\sum_{i}C_{\kappa i}^{L}g_{\kappa i}^{L}(r),\]

\begin{equation}
Q_{n\kappa}(r)=\sum_{i}C_{n\kappa}^{S}g_{\kappa i}^{S}(r),\label{gto}\end{equation}
where $S$ and $L$ stands for small and large component respectively
and the $g$'s are Gaussian type functions of the form

\begin{equation}
g_{\kappa i}^{L}(r)=C_{N}^{L}r^{n_{\kappa}}\exp(-\alpha_{i}r^{2}).\label{gto-2}\end{equation}
In the case of finite basis set expansions the condition of kinetic
balance is applied to prevent the variational collapse \cite{kinetic balance-lee}.
The kinetic balance condition \cite{stanton-jcp84,kin-bal-1,kin-bal-2}
gives the relation between the large and small component of radial
wave function as follows :

\begin{equation}
g_{\kappa i}^{S}(r)=C_{N}^{S}\left(\frac{d}{dr}+\frac{\kappa}{r}\right)g_{\kappa i}^{L}(r).\label{kb-1}\end{equation}
In Eq. (\ref{gto-2}) and (\ref{kb-1}) $C_{N}^{L}$ and $C_{N}^{S}$
are the normalization factors for the large and small components respectively.

\section{\label{beyond-DC}Beyond the Dirac-Coulomb Approximation}

The electron-electron interaction can be treated relativistically,
by including corrections to the Coulomb interaction. The leading relativistic
correction to the Coulomb interaction is the Breit interaction \cite{breit},
where the interaction Hamiltonian is given by, \begin{equation}
H_{B}=-e^{2}\sum_{i<j}\frac{\alpha_{i}\cdot\alpha_{j}}{r_{ij}}+\frac{\left(\alpha_{i}\cdot r_{ij}\right)\left(\alpha_{j}\cdot r_{ij}\right)}{r_{ij}^{3}}.\label{breit}\end{equation}
 Here the matrices $\alpha_{i},\alpha_{j}$ are built from the Dirac
matrices and $r_{ij}$ is the inter-electronic distance. The magnitude
of the Breit interaction is smaller than that of the Coulomb interaction
by a factor $\alpha^{2}$, where $\alpha$ is the fine structure constant
and it can be included perturbatively or self consistently \cite{quiney-qed}.
In addition to the Breit interaction, inclusion of QED effects like
the self-energy and the vacuum polarization \cite{vac-pol} may be
necessary for an accurate quantitative description of certain properties
where relativistic effects are important.

\begin{figure}
\begin{center}\includegraphics{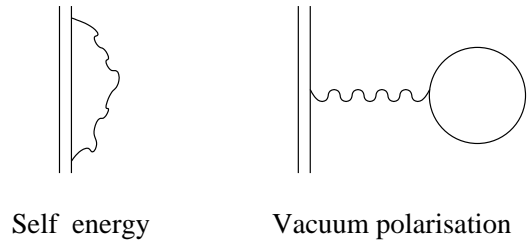}\end{center}

\caption{\label{qed}Corrections to Coulomb interaction}
\end{figure}

The process involving the emission and absorption of a virtual photon
by the same electron is known as self-energy. According to Dirac's
theory, the vacuum consists of a homogeneous sea of negative-energy
electrons. A bound electron in the atom can interact with an electron
in the Dirac sea, thereby changing the charge distribution of the
negative energy electrons compared to the free-field case. This results
in the creation of electron-positron pairs and hence the vacuum behaves
as a polarizable medium. This process is known as vacuum polarization.
Figure \ref{qed} illustrates the self-energy and the vacuum polarization
processes \cite{beier-phys-rep}.

Only a few calculations of self energy and the vacuum polarization
(which together give rise to the Lamb shift) have been performed on
many electron atoms. As the Coulomb interaction due to the nucleus
is much stronger than the electron-electron interactions in the inner
shells of heavy atoms, it is reasonable to calculate the Lamb shift
for such systems using the hydrogenic or screened hydrogenic approximation
\cite{desiderio-hydrogenic}. More sophisticated calculations of QED
effects have been carried out in the past few years. The details of
these calculations can be found in a review article by Shabaev \cite{shabaev}.

\section{\label{CC-Theory}Relativistic Coupled Cluster Theory : An overview}

In this section we will briefly introduce the relativistic coupled
cluster theory; one of the most powerful and accurate relativistic
many-body theories. It is equivalant to all order relativistic many-body
perturbation theory and has the virtue of being size-extensive \cite{size-extensive}.
Most of the work described in the subsequent section are based on
this theory.

We start with the DF state $\left|\Phi\right\rangle $ built out of
four component orbitals given by Eq. (\ref{dirac-spinor}), as the
Fermi vacuum, and then the normal ordered Hamiltonian can be expressed
as

\begin{equation}
H_{N}\equiv H-\left\langle \Phi\right|H\left|\Phi\right\rangle =H-E_{DF},\label{cc-1}\end{equation}
where $H$ is the Dirac-Coulomb Hamiltonian.

If we project $\left\langle \Phi\right|\exp(-T)$ from the left we
obtain the correlation energy ($\Delta E$) and if we project any
of the excited determinant $\left\langle \Phi^{\star}\right|\exp(-T)$
we additionally get a set of equations which are used to obtain the
$T$ amplitudes. Using the normal ordered dressed Hamiltonian $\overline{H}_{N}=\exp(-T)H_{N}\exp(T)$
the corresponding equations for correlation energy and amplitudes
become

\begin{equation}
\left\langle \Phi\right|\overline{H}_{N}\left|\Phi\right\rangle =\Delta E,\label{cc-2}\end{equation}
 and

\begin{equation}
\left\langle \Phi^{\star}\right|\overline{H}_{N}\left|\Phi\right\rangle =0.\label{cc-3}\end{equation}
 Here the state $\left|\Phi^{\star}\right\rangle $ may be singly
excited $\left|\Phi_{a}^{r}\right\rangle $ or double excited $\left|\Phi_{ab}^{rs}\right\rangle $
and so on. The indices $a,b,\cdots$ refer to holes and $p,q,\cdots$
to particles. We have considered the coupled cluster single and double
(CCSD) approximation, where the cluster operator $T$ is composed
of one- and two-body excitation operators, \emph{i.e.} $T=T_{1}+T_{2}$,
and are expressed in second quantization form

\begin{equation}
T=T_{1}+T_{2}=\sum_{ap}\left\{ a_{p}^{\dagger}a_{a}\right\} t_{a}^{p}+\frac{1}{2}\sum_{abpq}\left\{ a_{p}^{\dagger}a_{q}^{\dagger}a_{b}a_{a}\right\} t_{ab}^{pq}.\label{cc-4}\end{equation}

Contracting the ladder operators \cite{bartlett-book} and rearranging
the indices, the amplitude equations can be expressed in the form

\begin{equation}
A+B(T)\cdot T=0,\label{cc-5}\end{equation}
 where $A$ is a vector consisting of the matrix elements $\left\langle \Phi^{\star}\right|H_{N}\left|\Phi\right\rangle $
and $T$ is the vector representing the excitation amplitudes and
$B(T)$ is the matrix which depends on the cluster amplitudes so that
Eq. (\ref{cc-5}) is solved self-consistently. For example, a typical
contribution to the term $\overbrace{\overbrace{H_{N}T_{2}}T_{2}}$
is

\begin{equation}
B_{ab}^{pq}=\frac{1}{2}\sum_{dgrs}V_{dgrs}t_{ad}^{pr}t_{gb}^{sq}.\label{cc-6}\end{equation}
 Here $V_{dgrs}$ is the two-electron Coulomb integral and $t_{ad}^{pr}$
is the cluster amplitude corresponding to a simultaneous excitation
of two electrons from orbital $a$ and $d$ to $p$ and $r$ respectively.
Diagrammatic techniques are used to obtain all the terms which contribute
to this specific contribution.

For an atom with one valance electron we first compute the correlations
for the closed shell system, \emph{i.e.} singly ionized atom using
the closed shell coupled cluster approach. The reference state for
the open shell system is

\begin{equation}
\left|\Phi_{k}^{N+1}\right\rangle \equiv a_{k}^{\dagger}\left|\Phi\right\rangle \label{cc-7}\end{equation}
 with the particle creation operator $a_{k}^{\dagger}$. Then by using
the excitation operators for both the core and valance electron the
exact state is defined as \cite{lindgren-book}:

\begin{equation}
\left|\Psi_{k}^{N+1}\right\rangle =\exp(T)\left\{ \exp(S_{k})\right\} \left|\Phi_{k}^{N+1}\right\rangle .\label{cc-8}\end{equation}
 Here $\left\{ \exp(S_{k})\right\} $ is the normal ordered exponential
representing the valance part of the wave operator. Here

\begin{equation}
S_{k}=S_{1k}+S_{2k}=\sum_{k\neq p}\left\{ a_{p}^{\dagger}a_{k}\right\} s_{k}^{p}+\sum_{bpq}\left\{ a_{p}^{\dagger}a_{q}^{\dagger}a_{b}a_{k}\right\} s_{kb}^{pq}\,,\label{cc-9}\end{equation}
 where $k$ stands for valance orbital. $S_{k}$ contain the particle
annihilation operator $a_{k}$, and because of the normal ordering
it cannot be connected to any other valance electron excitation operator
and so $\left\{ \exp(S_{k})\right\} $ automatically reduces to $\left(1+S_{k}\right)$.

Then we can write the Eq.(\ref{cc-8}) as

\begin{equation}
\left|\Psi_{k}^{N+1}\right\rangle =\exp(T)\left(1+S_{k}\right)\left|\Phi_{k}^{N+1}\right\rangle ,\label{cc-10}\end{equation}
 and obtain a set of equations \cite{bartlett-book}

\begin{equation}
\left\langle \Phi_{k}^{N+1}\right|\overline{H}_{N}\left(1+S_{k}\right)\left|\Phi_{k}^{N+1}\right\rangle =H_{eff}\label{cc-11}\end{equation}
 and

\begin{equation}
\left\langle \Phi_{k}^{^{\star}N+1}\right|\overline{H}_{N}\left(1+S_{k}\right)\left|\Phi_{k}^{N+1}\right\rangle =H_{eff}\left\langle \Phi_{k}^{^{\star}N+1}\right|\left(1+S_{k}\right)\left|\Phi_{k}^{N+1}\right\rangle .\label{cc-12}\end{equation}
 The Eq.(\ref{cc-12}) is non-linear in $S_{k}$ because $H_{eff}$
is itself a function of $S_{k}$. Hence, these equations have to solved
self-consistently to determine the $S_{k}$ amplitudes.

The normalized transition matrix element ($i\longrightarrow f$) due
to an operator $\widehat{O}$ is given by

\begin{equation}
\begin{array}{cc}
\widehat{O}_{fi} & =\frac{\left\langle \Psi_{f}^{N+1}\right|\widehat{O}\left|\Psi_{i}^{N+1}\right\rangle }{\sqrt{\left\langle \Psi_{f}^{N+1}\right|\left.\Psi_{f}^{N+1}\right\rangle \left\langle \Psi_{i}^{N+1}\right|\left.\Psi_{i}^{N+1}\right\rangle }}\\
 & =\frac{\left\langle \Phi_{f}^{N+1}\right|\left\{ 1+S_{f}^{\dagger}\right\} \exp(T^{\dagger})\widehat{O}\exp(T)\left\{ 1+S_{i}\right\} \left|\Phi_{i}^{N+1}\right\rangle }{\sqrt{\left\langle \Phi_{f}^{N+1}\right|\left\{ 1+S_{f}^{\dagger}\right\} \exp(T^{\dagger})\exp(T)\left\{ 1+S_{f}\right\} \left|\Phi_{f}^{N+1}\right\rangle \left\langle \Phi_{i}^{N+1}\right|\left\{ 1+S_{i}^{\dagger}\right\} \exp(T^{\dagger})\exp(T)\left\{ 1+S_{i}\right\} \left|\Phi_{i}^{N+1}\right\rangle }}\,,\end{array}\label{tr-mat}\end{equation}
 whereas the expectation value of any operator $\widehat{O}$ can
be written as the normalized form with respect to the exact state
$\left|\Psi^{N+1}\right\rangle $ as

\begin{equation}
\left\langle \widehat{O}\right\rangle =\frac{\left\langle \Psi^{N+1}\right|\widehat{O}\left|\Psi^{N+1}\right\rangle }{\left\langle \Psi^{N+1}\right|\left.\Psi^{N+1}\right\rangle }=\frac{\left\langle \Phi^{N+1}\right|\left\{ 1+S^{\dagger}\right\} \exp(T^{\dagger})\widehat{O}\exp(T)\left\{ 1+S\right\} \left|\Phi^{N+1}\right\rangle }{\left\langle \Phi^{N+1}\right|\left\{ 1+S^{\dagger}\right\} \exp(T^{\dagger})\exp(T)\left\{ 1+S\right\} \left|\Phi^{N+1}\right\rangle }.\label{exp-val}\end{equation}

\section{\label{Rel-Eff}Purely relativistic effects}

\subsection{\label{fine-str.}Fine-Structure splitting}

The fine-structure splitting is relativistic in origin, but is influenced
by electron correlation. It occurs between the states having same
values of the total orbital quantum number $L$, total spin quantum
number $S$ and different total angular momentum $J$. There have
been many attempts to calculate this quantity for a variety of atoms
in their ground and excited states \cite{pyper}. We present here
the interesting case of the ground state fine structure splitting
of boron which has been calculated by different relativistic approaches
\cite{froese-fischer,frye,bpdas-boron}. Boron is an open-shell atom
with the configuration $1s^{2}2s^{2}2p^{1}$. The relativistic configuration
interaction method (CI) was used by Das \emph{et al} \cite{bpdas-boron}.
The single particle orbitals used in the calculations there were obtained
by the application of the variational principle. Consider an energy
functional given by \begin{equation}
\mathbf{\varepsilon}=\sum_{r}a_{r}\langle\Phi_{r}|H|\Phi_{r}\rangle\label{en-func}\end{equation}
 where $H$ is the Dirac-Coulomb Hamiltonian, $|\Phi_{r}\rangle$is
the $r$th configuration state function (CSF) and $a_{r}$ is given
by \begin{equation}
a_{r}=\frac{2J_{r}+1}{\sum_{s}(2J_{s}+1)}\,;\label{a_r}\end{equation}
 $J_{r}$ and $J_{s}$ being the total angular momenta of the $r$th
and $s$th CSFs respectively. Minimization of $\varepsilon$ with
respect to the single particle orbitals $\phi_{i}$, \begin{equation}
\frac{\partial\varepsilon}{\partial\phi_{i}}=0\label{variation}\end{equation}
 yields a set of differential equations which were solved self-consistently
by using an appropriate numerical method \cite{grant-adv}. In this
calculation, all relativistic configurations arising from $1s^{2}2s^{2}2p^{1}$,
$1s^{2}2s^{2}2p^{2}$and $1s^{2}2p^{3}$ were considered. The Breit
interaction and QED effects (self-energy and vacuum polarization in
the hydrogenic approximation) were treated as first order perturbations.
We give below the results of the calculation. The electron correlation
contributions (difference of Dirac-Coulomb+Breit and Dirac-Fock) vary
from $-1078.8$ to $-5287.4$ for $Z=20$ to $Z=30$. It is the evident
from table.\ref{fs} that the Breit interaction and the QED effects
play an important role and their inclusion is critical in obtaining
good agreement with experiments.

\begin{table}

\caption{\label{fs}Fine structure intervals for B-like ions in $cm^{-1}$. }

\begin{center}\begin{tabular}{ccccccc}
\hline 
Z &
 Dirac-Fock (DF)&
 Dirac-Coulomb (DC) &
 Breit &
 QED &
 Total &
 Experiment \tabularnewline
\hline
\hline 
20 &
 37581.8&
 38119 &
 -1616 &
 124 &
 36627 &
 36615(30) \tabularnewline
22 &
 57678.6&
 58319 &
 -2241 &
 180 &
 56258 &
 56243(4) \tabularnewline
24 &
 84973.2&
 85715 &
 -3015 &
 251 &
 82951 &
 829926(20) \tabularnewline
26 &
 121045.8&
 121898 &
 -3958 &
 341 &
 118281 &
 118266(20) \tabularnewline
28 &
 167653.8&
 168635 &
 -5092 &
 451 &
 163994 &
 163961(50) \tabularnewline
30 &
 226739.4&
 227889 &
 -6437 &
 586 &
 222038 &
\tabularnewline
\hline
\hline 
&
&
&
&
&
&
\\
\tabularnewline
\end{tabular}\end{center}
\end{table}

\subsection{\label{edm}Electric Dipole Moment of the electron}

The presence of a non-zero electric dipole moment (EDM) on a non-degenerate
physical system would be a direct evidence of Parity ($\hat{P}$)
and Time-reversal ($\hat{T}$) symmetry violations. An atom can have
a non-zero EDM due to a non-zero EDM of it's constituent electron,
under certain conditions. According to a theorem by Schiff, in 1963
\cite{schiff}, \emph{the EDM of an atom vanishes even if it's constituents
have non-vanishing EDMs}. This theorem was based on the following
assumptions :

\begin{enumerate}
\item the constituents of the atoms are non-relativistic particles, 
\item the interactions between the particles in an atom are electrostatic, 
\item the EDM distribution of each atomic constituent coincides with its
charge distribution. 
\end{enumerate}
By considering the relativistic effects in atoms, Sandars showed that
an atom can have a non-zero EDM \cite{sanders}. If an electron has
a non-zero EDM $d_{e}$, the relativistic interaction of $d_{e}$
with the internal electric field of the atom, is given by, \begin{equation}
H_{I}=-\sum_{i}d_{e}\beta_{i}\vec{\sigma}_{i}\cdot\vec{E}_{i}^{int},\label{edm-1}\end{equation}
 where $E^{int}$ is the electric filed inside the atom; $\vec{\sigma}_{i}$
are Pauli matrices and $\beta$ is Dirac matrix defined in section
\ref{dirac-coul}. This reduces to, \begin{equation}
H_{I}=-\sum_{i}d_{e}\vec{\sigma}_{i}\cdot\vec{E}_{i}^{int}\label{edm-2}\end{equation}
 in the non-relativistic limit. It is possible to express the relativistic
form of $H_{I}$ in terms of an effective one particle Hamiltonian,
given by \cite{shukla}\begin{equation}
H_{I}=2ic\beta_{i}\gamma_{i}^{5}p_{i}^{2}\,,\label{edm3}\end{equation}
 where $c$ is the velocity of light, $\gamma_{5}=i\gamma_{0}\gamma_{1}\gamma_{2}\gamma_{3}$
and $\gamma_{i}=\beta\alpha_{i}$. The Schr$\mathrm{\ddot{o}}$dinger
equation for the unperturbed state $|\Psi_{\alpha}^{(0)}\rangle$
is

\begin{equation}
H_{0}|\Psi_{\alpha}^{(0)}\rangle=E_{\alpha}^{(0)}|\Psi_{\alpha}^{(0)}\rangle\,,\label{unp-ev-eqn}\end{equation}
 where $|\Psi_{\alpha}^{(0)}\rangle=\exp(T^{(0)})|\Phi_{\alpha}^{(0)}\rangle$
in coupled-cluster theory and $H_{0}$ is the Dirac-Coulomb Hamiltonian.

In the presence of EDM interaction, which is treated as a perturbation,
the Schr$\ddot{o}$dinger equation becomes

\begin{equation}
H|\Psi_{\alpha}\rangle=E_{\alpha}|\Psi_{\alpha}\rangle\,,\label{per-ev-eqn}\end{equation}
 where $H=H_{0}+\lambda H_{I}$ and $|\Psi_{\alpha}\rangle=\exp(T^{(0)}+\lambda T^{(1)})|\Phi_{\alpha}^{(0)}\rangle$.
Here $T^{(0)}$and $T^{(1)}$are the unperturbed and perturbed cluster
amplitudes and the perturbation parameter, $\lambda=d_{e}$. The $T^{(0)}$and
$T^{(1)}$amplitudes are determined from the following equations :

\begin{equation}
\left\langle \Phi^{\star}\right|\overline{H}_{N}\left|\Phi\right\rangle =0\label{t0-eqn}\end{equation}
and

\begin{equation}
\left\langle \Phi^{\star}\right|\left[\overline{H}_{N},T^{(1)}\right]|\Phi_{\alpha}^{(0)}\rangle=-\left\langle \Phi^{\star}\right|\overline{H_{I}}|\Phi_{\alpha}^{(0)}\rangle.\label{t1-eqn}\end{equation}
 The atomic EDM is given by, \begin{equation}
d_{A}=\frac{\langle\Psi_{\alpha}|\vec{D}|\Psi_{\alpha}\rangle}{\langle\Psi_{\alpha}|\Psi_{\alpha}\rangle}=0\label{edm-5}\end{equation}
 for non-relativistic case. $\vec{D}$ is the electric dipole operator.
The enhancement factor $R$ is given by, \begin{equation}
R=\frac{d_{a}}{d_{e}}=\frac{\langle\Psi_{\alpha}^{(0)}|\vec{D}|\Psi_{\alpha}^{(1)}\rangle+\langle\Psi_{\alpha}^{(1)}|\vec{D}|\Psi_{\alpha}^{(0)}\rangle}{\langle\Psi_{\alpha}^{(0)}|\Psi_{\alpha}^{(0)}\rangle}\label{edm-6}\end{equation}
 Following Coupled Cluster theory the equation (Eq. \ref{edm-6})
reduces to\begin{equation}
R=\frac{\langle\Phi_{0}|T^{(1)^{\dagger}}\vec{D}+\vec{D}T^{(1)}|\Phi_{0}\rangle}{\langle\Phi_{0}|\Phi_{0}\rangle}\,.\label{edm-8}\end{equation}
 An alternative Coupled Cluster approach to EDMs is given by Shukla
et al \cite{shukla-das-dm}. The values of $d_{e}$ predicted by various
models of particle physics are given in table \ref{diff-de}.

\begin{table}

\caption{\label{diff-de}Value of $d_{e}$ predicted by various models of
particle physics}

\begin{center}\begin{tabular}{cc}
\hline 
Model &
 $d_{e}$in $e-cm$\tabularnewline
\hline
\hline 
Standard Model &
 $<10^{-38}$\tabularnewline
Supersymmetric &
 $10^{-26}-10^{-28}$\tabularnewline
Multi-Higgs &
 $10^{-26}-10^{-28}$\tabularnewline
Left-right asymmetric &
 $10^{-26}-10^{-28}$\tabularnewline
\hline
\hline 
&
\\
\tabularnewline
\end{tabular}\end{center}
\end{table}

The current best limit on the electron EDM comes from the $Tl$ measurement
\cite{kuvichev}. The enhancement factor atomic thallium to the electron
EDM $R=-585$, which is based on relativistic coupled-cluster calculation
\cite{liu}. Comparing with experiment, the limit on the electron
EDM is\begin{equation}
d_{e}\leq1.6\times10^{-27}\mbox{e-cm}.\label{edm-9}\end{equation}
 The enhancement factor for atomic $Cs$ (Z=55) has been obtained
as $R=130.5$ from a method combining RMBPT and the MCDF approach
\cite{das-lec-notes}. The calculation done by Martensson \emph{et
al} \cite{martensson} gives $R=114(1\pm0.03)$ for $Cs$.

\section{\label{rel-enhance}Relativistic enhancements}

\subsection{\label{corr-eff}Correlation energy}

In the frame work of coupled-cluster theory, the expression for the
correlation energy of an atom is given by, \begin{equation}
E_{corr}=\langle\Phi|\bar{H}_{N}|\Phi\rangle,\label{corr-energy}\end{equation}
 where $\bar{H}_{N}=e^{-T}H_{N}e^{T}$ where $H$ is the Dirac-Coulomb
Hamiltonian described in section \ref{dirac-coul}.

\begin{figure}
\begin{center}\includegraphics{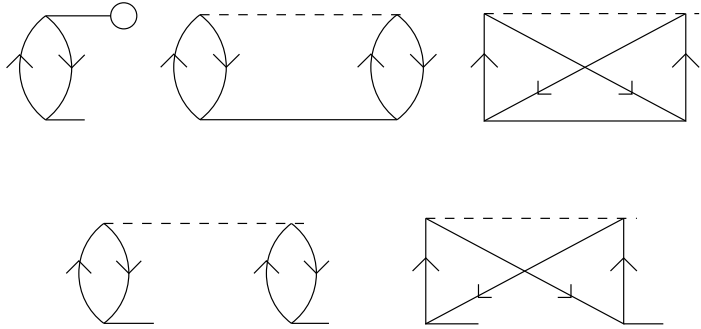}\end{center}

\caption{\label{corr-en-diag}Diagrams contributing to correlation energy}
\end{figure}

The diagrams given in figure \ref{corr-en-diag} contribute to the
correlation energy, where the dotted lines represent the coulomb interaction
between the electrons, the solid line corresponds to the cluster operator
$T$ and the circle represents the one-electron operator.

We have computed the above expression for the correlation energy using
the coupled-cluster wavefunctions for the closed-shell atoms, $Xe^{54}$,
$Yb^{70}$ and $Hg^{80}$. The results are shown in table \ref{corr-en}.

\begin{table}

\caption{\label{corr-en}Comparison of correlation and Dirac-Fock energy}

\begin{center}\begin{tabular}{ccc}
\hline 
Atom &
 Dirac Fock energy &
 $\delta E_{corr}$\tabularnewline
\hline
\hline 
$Xe^{54}$&
 -0.74474960061E+04 &
 -0.71694286411 \tabularnewline
$Yb^{70}$&
 -0.14069217432E+05 &
 -0.56956394691 \tabularnewline
$Hg^{80}$&
 -0.19650686115E+05 &
 -0.44792843655 \tabularnewline
\hline
\hline 
&
&
\tabularnewline
\end{tabular}\end{center}
\end{table}

From the above results it is clear that with the increase in the atomic
number ($Z$), the relativistic effects become more prominent. The
absolute magnitude of the Dirac-Fock contribution hence increases
and that of the correlation energy decreases for xenon, ytterbium
and mercury. 

The orbitals used in the calculation are expanded in terms of Gaussian
functions of the type \cite{napp-fbse}

\begin{equation}
F_{i,k}(r)=r^{k}\exp(-\alpha_{i}r^{2}),\label{comp-1}\end{equation}
with $k=0,1,2\cdots$ for $s,p,d,\cdots$ type functions, respectively.
The exponents are determined by the even tempering condition \cite{even-temp}

\begin{equation}
\alpha_{i}=\alpha_{0}\beta^{i-1}.\label{comp-2}\end{equation}
The values of $\alpha_{0}$and $\beta$ for different symmetries are
given in table \ref{tab3}.

\begin{table}

\caption{\label{tab3}Details of the basis used in the calculation}

\begin{center}\begin{tabular}{cccc}
\hline 
symmetry &
 Total basis in &
 No. of excited &
 $\alpha_{0}$ and $\beta$ used \tabularnewline
&
 each symmetry &
 orbitals &
\tabularnewline
\hline
\hline 
$s_{1/2}$&
 13 &
 8 &
 0.00725 ; 2.725 \tabularnewline
$p_{1/2}$&
 11 &
 7 &
 0.00755 ; 2.755 \tabularnewline
$p_{3/2}$&
 11 &
 7 &
 0.00755 ; 2.755 \tabularnewline
$d_{3/2}$&
 8 &
 2 &
 0.00775 ; 2.765 \tabularnewline
$d_{5/2}$&
 8 &
 2 &
 0.00775 ; 2.765 \tabularnewline
$f_{5/2}$&
 5 &
 5 &
 0.00780 ; 2.805 \tabularnewline
$f_{7/2}$&
 5 &
 5 &
 0.00780 ; 2.805 \tabularnewline
$g_{7/2}$&
 3 &
 3 &
 0.00785 ; 2.825 \tabularnewline
$g_{9/2}$&
 3 &
 3 &
 0.00785 ; 2.825 \tabularnewline
\hline
\hline 
&
&
&
\\
\tabularnewline
\end{tabular}\end{center}
\end{table}

\subsection{\label{hyperfine}Hyperfine interaction}

A nucleus may possess electromagnetic multipole moments, which can
interact with the electromagnetic field produced by the electrons
at the site of the nucleus . The interaction between various moments
of the nucleus and the electrons of an atom are collectively known
as hyperfine interactions \cite{lindgren-book}. This interaction
produce shifts of the electronic energy levels which are usually much
smaller than those corresponding to the fine structure splittings.

The non-vanishing moments are the magnetic multipole moments for odd
$k$ and electric multipole moments for even $k$. The most important
of these moments is the magnetic dipole moment ($k=1$) which is associated
with the nuclear spin. The interaction of this particular moment with
the electron is known as magnetic dipole hyperfine interaction.

In general the hyperfine interaction is given by \cite{cheng}

\begin{equation}
H_{hfs}=\sum_{k}M^{(k)}\cdot T^{(k)},\label{eqn-1}\end{equation}
 where $M^{(k)}$ and $T^{(k)}$ are spherical tensors of rank $k$,
which corresponds to nuclear and electronic parts of the interaction
respectively.\\
 For the magnetic dipole hyperfine interaction \cite{lind-case-st}

\begin{equation}
T_{q}^{(1)}=\sum_{q}t_{q}^{(1)}=\sum_{j}-ie\sqrt{\frac{8\pi}{3}}\frac{\overrightarrow{\alpha_{j}}}{r_{j}^{2}}\cdot\mathbf{Y}_{1q}^{(0)}(\widehat{r_{j}}),\label{eqn-4}\end{equation}
 where $\overrightarrow{\alpha}$ is the Dirac matrix and $\mathbf{Y}_{kq}^{\lambda}$
is the vector spherical harmonics. In Eq.(\ref{eqn-4}) the index
$j$ refers to the $j$-th electron of the atom and $e$ is the magnitude
of the electronic charge. The magnetic dipole hyperfine constant $A$
is defined as

\begin{equation}
A=\mu_{N}\left(\frac{\mu_{I}}{I}\right)\frac{\left\langle J\right\Vert T^{(1)}\left\Vert J\right\rangle }{\sqrt{J(J+1)(2J+1)}},\label{eqn-5}\end{equation}
 where $\mu_{N}$ is the nuclear Bohr magneton, $\mu_{I}$ is the
nuclear magnetic moment, $I$ is the nuclear spin, $J$ is the total
electronic angular momentum.

In Eq.(\ref{eqn-4}) $t^{(1)}$ is the single particle reduced matrix
element of $T^{(1)}$. The reductions of the single particle matrix
element into angular factors and radial integral can be obtained by
using the Wigner Eckart theorem. This single particle reduced matrix
element is given by

\begin{equation}
\left\langle \kappa\right\Vert t^{(1)}\left\Vert \kappa^{\prime}\right\rangle =-\left\langle \kappa\right\Vert C^{(1)}\left\Vert \kappa^{\prime}\right\rangle (\kappa+\kappa^{\prime})\int dr\frac{\left(P_{\kappa}Q_{\kappa^{\prime}}+Q_{\kappa}P_{\kappa^{\prime}}\right)}{r^{2}}\,,\label{eqn-7}\end{equation}
 where $\left\langle \kappa\right\Vert C^{(k)}\left\Vert \kappa^{\prime}\right\rangle $
is the reduced matrix element of the Racah tensor and is equal to

\[
(-1)^{j+1/2}\sqrt{(2j+1)(2j^{\prime}+1)}\left(\begin{array}{ccc}
j & k & j^{\prime}\\
\frac{1}{2} & 0 & -\frac{1}{2}\end{array}\right)\pi(l,k,l^{\prime}),\]
 with

\[
\pi(l,k,l^{\prime})=\left\{ \begin{array}{c}
\begin{array}{cc}
1 & \mathrm{if}\: l+k+l^{\prime}\,\,\mathrm{even}\\
0 & \mathrm{otherwise}\end{array}\end{array}\right..\]
 Here the single particle orbitals are expressed in terms of the Dirac
spinors with $P_{i}$ and $Q_{i}$ as large and small components respectively.

In the calculation for $Ba^{+}$we have used hybrid basis functions
which are partly numerical and partly analytical \cite{hybrid}. The
analytical orbitals have the form of Eq. (\ref{comp-1}). In table
\ref{hypA} the values of the magnetic dipole hyperfine constant ($A$)
is given in MHz for $^{25}Mg^{+}$\cite{csur-mg+} and $^{137}Ba^{+}$
\cite{napp-hyp-ba+} for ground and one low lying excited state. In
table \ref{cp-pc cont} we have presented the contributions from Dirac-Fock
(DF), pair correlation (PC) and core polarization (PC) effects. 

\begin{figure}
\begin{center}\includegraphics{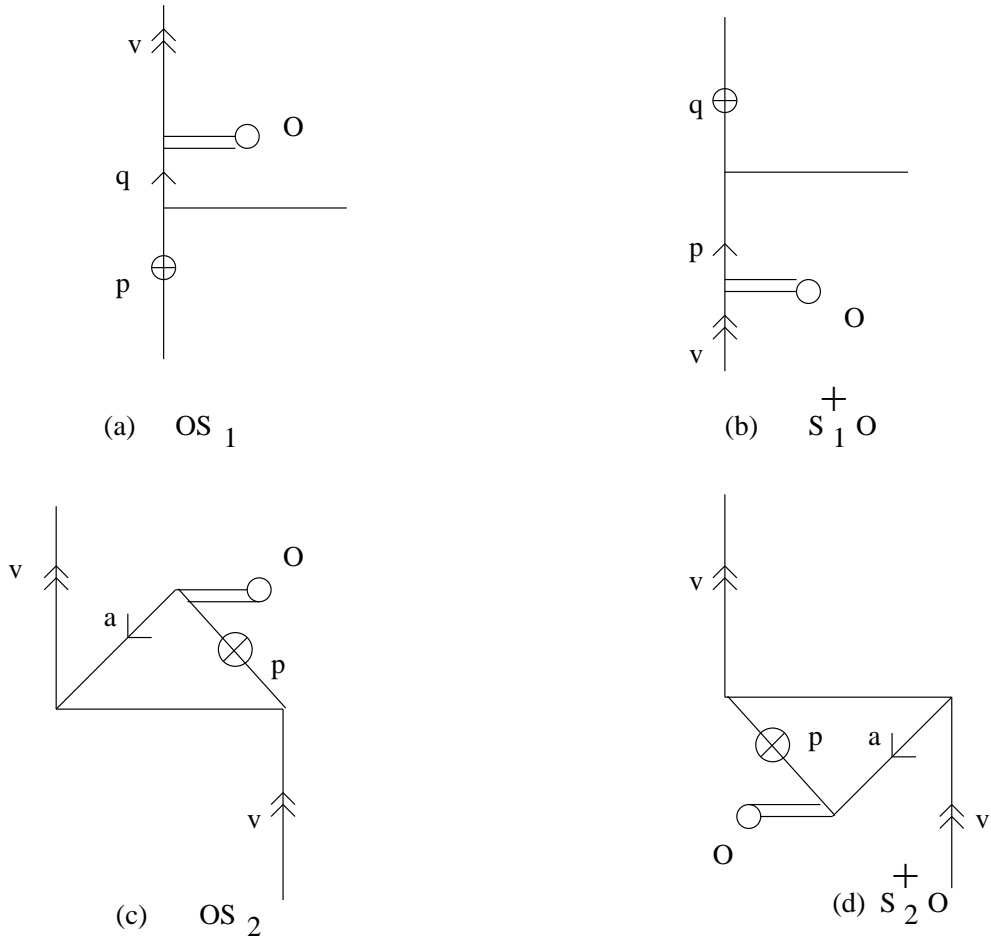}\end{center}

\caption{\label{diag-cc}Goldstone diagrams for pair correlation (a,b) and
core-polarization effects (c,d). Here $a$ denotes a hole whereas
$v$ denotes valance orbital and $p,q,r...$ denote virtual orbitals
(particles). The superscripts refer to the order of perturbation and
the dashed lines correspond to the Coulomb interaction. Particles
and holes (labeled by $a$) are denoted by the lines directed upward
and downward respectively. The double line represents the O (the hyperfine
interaction operator) vertices. The valance (labeled by $v$) and
virtual orbitals (labeled by $p,q,r..$) are depicted by double arrow
and single arrow respectively, whereas the orbitals denoted by $\oplus$
can either be valance or virtual.}
\end{figure}

\begin{table}

\caption{\label{hypA}Values of magnetic dipole hyperfine constant (A) in
MHz for $^{25}Mg^{+}$and $^{137}Ba^{+}$}

\begin{center}\begin{tabular}{lllll}
\hline 
Atom&
&
 Theory&
 Experiment&
 Others \tabularnewline
\hline
\hline 
&
&
&
&
\tabularnewline
$^{25}Mg^{+}$&
 States&
&
&
\tabularnewline
&
 $3s_{1/2}$&
 592.86&
 596.25&
 602(8) \cite{tpdas-mg+}\tabularnewline
&
&
&
&
 597.45 \cite{johnson-hyp}\tabularnewline
&
 $3p_{1/2}$&
 101.70&
&
 103.4 \cite{johnson-hyp}\tabularnewline
&
&
&
&
\tabularnewline
$^{137}Ba^{+}$&
&
&
&
\tabularnewline
&
 $6s_{1/2}$&
 4072.83&
 4018&
 4203.200 \cite{tpdas-ba+}\tabularnewline
&
 $6p_{1/2}$&
 736.98&
 742.04&
\tabularnewline
&
&
&
&
\tabularnewline
\hline
\hline 
&
&
&
&
\\
\tabularnewline
\end{tabular}\end{center}
\end{table}

It can be seen from this table (table \ref{cp-pc cont}) that for
$Mg^{+}$ the CP contribution is larger than the PC in magnitude for
both the states. It is important to note that the former contribution
includes the hyperfine interaction of all the core orbitals while
only a specific valence orbital is involved in this interaction for
the latter (see Fig. \ref{diag-cc}). However, the hyperfine constant
$A$ for $Ba^{+}$ exhibits exactly the opposite behaviour. Even though
$Ba^{+}$ has more core electrons than $Mg^{+}$, the relativistic
enhancement of the valence ($6s$) magnetic dipole hyperfine interaction
results in the value of PC exceeding that of CP.

\begin{table}

\caption{\label{cp-pc cont}Contribution of pair correlation (PC) and core
polarization (CP) effect in magnetic dipole hyperfine constant (A)
in MHz}

\begin{center}\begin{tabular}{cllll}
\hline 
Atom&
&
 Dirac-Fock&
 PC&
CP\tabularnewline
\hline
\hline 
&
&
&
&
\tabularnewline
$^{25}Mg^{+}$\cite{csur-mg+}&
 States&
&
&
\tabularnewline
&
 $3s_{1/2}$&
 468.819&
 39.713&
 77.767\tabularnewline
&
 $3p_{1/2}$&
 77.975&
 7.293&
 15.153\tabularnewline
&
&
&
&
\tabularnewline
$^{137}Ba^{+}$\cite{napp-hyp-ba+}&
&
&
&
\tabularnewline
&
 $6s_{1/2}$&
 2929.41&
 663.20&
 465.91\tabularnewline
&
 $6p_{1/2}$&
 492.74&
 126.53&
 98.98\tabularnewline
&
&
&
&
\tabularnewline
\hline
\hline 
&
&
&
&
\tabularnewline
\end{tabular}\end{center}
\end{table}

\begin{table}

\caption{\label{pb+A}Magnetic dipole hyperfine constant for $6p$ states
of $Pb^{+}$: a strongly interacting system. RMBPT(2) stands for second
order RMBPT. Both RMBPT and RCCSD calculations are performed by our
group.}

\begin{center}\begin{tabular}{lll}
\hline 
States&
 $6p\,^{2}P_{1/2}$&
 $6p\,^{2}P_{3/2}$\tabularnewline
\hline
\hline 
&
&
\tabularnewline
Dirac-Fock&
 11513.5&
 918.4\tabularnewline
RMBPT(2)&
 15722.5&
 302.9\tabularnewline
RCCSD(T)&
 12903.7&
 623.2\tabularnewline
Experiment \cite{pb+A-exp}&
 13000&
 583(21)\tabularnewline
&
&
\tabularnewline
\hline
\hline 
&
&
\tabularnewline
\end{tabular}\end{center}
\end{table}

In table \ref{pb+A} we present the values of $A$ for $6p\,^{2}P_{1/2}$
and $6p\,^{2}P_{3/2}$ states of $Pb^{+}$. The Dirac-Fock values
for these two states deviate from their experimental values in opposite
direction, suggesting that the signs of the correlation contributions
are opposite for the two cases. This is evident from the result of
our second order relativistic many-body perturbation theory (RMBPT(2))
calculation. Electron correlation is dramatic in the case of the $6p\,^{2}P_{3/2}$
state because of the large and negative core polarization (-840.6
MHz). However, the value of $A$ at this level differs from experiment
by 48\%. After carrying out a RCCSD(T) calculation this discrepancy
reduces to less than 7\% . The agreement of the ground state value
of $A$ with experiment is about 0.7\%. These calculations highlight
the power of the relativistic coupled-cluster theory to account for
the interplay of relativistic and correlation effects in systems with
strongly interacting configurations \cite{sahoo-pb+}.

\subsection{\label{PNC}Parity non-conservation in atoms to neutral weak interaction}

The parity transformation can be expressed as $\overrightarrow{r}\longrightarrow-\overrightarrow{r}$
and the action of the parity operator $\hat{P}$ is given by

\begin{equation}
\hat{P}\psi(\overrightarrow{r})=\psi(-\overrightarrow{r})\,,\label{pnc-1}\end{equation}
 where $\psi(\overrightarrow{r})$ is the wavefunction of a physical
system. Parity conservation means that the system is invariant under
parity transformation and the Hamiltonian $H$ commutes with the parity
operator, i.e. if $H_{P}$ is the parity transformed Hamiltonian

\[
H_{p}=\hat{P}H\hat{P}^{-1}=H\]
 and therefore

\begin{equation}
\left[H,\hat{P}\right]=0\,.\label{pnc-2}\end{equation}
 Hence it clearly implies that parity non-conservation (parity violation)
means that its Hamiltonian does not commute with the parity operator
$\hat{P}$.

Parity non-conservation (PNC) was discovered in the beta decay of
$^{60}Co$ by Wu and co-workers in 1957 following the prediction by
Lee and Yang a year earlier \cite{cswu-1957}. This lack of mirror
symmetry has now been observed in several systems and even in atoms
which is an important phenomenon to study. The latest measurement
of parity non-conservation in cesium with unprecedented accuracy (0.35\%)
has led to the discovery of the nuclear anapole moment \cite{nuc-anapole}.

The dominant contribution to PNC in atoms comes from the neutral weak
current (NWC) interaction between the electron and the nucleus \cite{nwc-book}.
The effective Hamiltonian describing the interaction consists of two
parts, one of which is nuclear spin independent (NSI) \cite{nwc-book}
and the other is nuclear spin dependent (NSD) \cite{flambaum-nsd,bouchita-nsd}.
In this review article we will concentrate on NSI parity non-conservation
in atoms. The NSI effective Hamiltonian is expressed as

\begin{equation}
H_{PNC}=\frac{G_{F}}{2\sqrt{2}}Q_{W}\sum_{e}\gamma_{5}^{e}\rho(r_{e}),\label{pnc-3}\end{equation}
 with

\begin{equation}
Q_{W}=2\left[ZC_{1p}+NC_{1n}\right].\label{pnc-4}\end{equation}
 Here $Z$ and $N$ are the number of protons and neutrons respectively
and $C_{1p}$ and $C_{1n}$ are the vector (nucleon) - axial vector
(electron) coupling coefficients whereas $G_{F}$ is the Fermi coupling
constant and $\rho(r_{e})$ is the normalized nucleon number density.
The matrix element of $H_{PNC}$ scales as $Z^{3}$ and it has been
treated as a perturbation. It is primarily because of this reason
that the heavy atoms are considered to be the best candidates for
PNC experiments. The total Hamiltonian is now represented by

\begin{equation}
H=H_{0}+H_{PNC}.\label{hamiltonian}\end{equation}
 This perturbation causes the wavefunction to take the form $\left|\Psi\right\rangle =\left|\Psi^{(0)}\right\rangle +\left|\Psi^{(1)}\right\rangle $,
where $\left|\Psi^{(0)}\right\rangle $ and $\left|\Psi^{(1)}\right\rangle $
are the unperturbed and the perturbed part of the wave function respectively.

The quantity that is measured in such an experiment depends on the
interference of a parity non-conserving electric dipole transition
amplitude ($E1_{PNC}$) and an allowed transition amplitude corresponding
to two atomic states of the same parity \cite{e1-all-tran}. From
the theoretical point of view an accurate calculation of $E1_{PNC}$
must be based on a suitable and accurate relativistic many-body theory.
In a recent review, Ginges and Flambaum \cite{pnc-phys-rep} have
presented the current status of atomic PNC calculations and experiments.
A number of many-body theories have been applied to calculate $E1_{PNC}$
matrix elements. The results of these calculations in combination
with the most accurate PNC experiment on $Cs$ is in agreement with
the Standard Model (SM) of particle physics \cite{pnc-phys-rep}. 

We have formulated a new approach to PNC in atoms based on relativistic
CC theory in an attempt to go beyond the existing calculations. In
this formulation the excitation operators (both $T$ and $S$) contain
an unperturbed (superscript $0$) and a perturbed part (superscript
$1$). For a single valence systems like cesium the wavefunction can
be written as 

\begin{equation}
\left|\Psi_{k}\right\rangle =\exp(T^{0}+T^{(1)})\left\{ 1+S^{(0)}+S^{(1)}\right\} \left|\Phi_{k}\right\rangle .\label{pnc-wfn}\end{equation}
 This equation follows from Eq.(\ref{cc-10}) and can be derived easily
\cite{geetha-th}. The equations for determining $T^{(1)}$and $S^{(1)}$
amplitudes are the following :

\begin{equation}
\left\langle \Phi^{\star}\right|\left[\overline{H}_{N},T^{(1)}\right]\left|\Phi_{0}\right\rangle +\left\langle \Phi^{\star}\right|\overline{H}_{PNC}\left|\Phi_{0}\right\rangle =0\,,\label{t1-eqn}\end{equation}
and

\begin{equation}
\left\langle \Phi_{v}^{k}\right|\overline{H}_{N}S_{v}^{(1)}-\Delta E_{v}^{(0)}S_{v}^{(1)}\left|\Phi_{v}\right\rangle +\left\langle \Phi_{v}^{k}\right|\overline{H}_{N}\left\{ T^{(1)}+T^{(1)}S_{v}^{(0)}\right\} +\overline{H}_{PNC}\left\{ 1+S_{v}^{(0)}\right\} \left|\Phi_{0}\right\rangle =0\,.\label{s1-eqn}\end{equation}

The parity non-conserving electric dipole transition amplitude between
atomic states $\left|\Psi_{i}\right\rangle =\left|\Psi_{i}^{(0)}\right\rangle +\left|\Psi_{i}^{(1)}\right\rangle $
and $\left|\Psi_{f}\right\rangle =\left|\Psi_{f}^{(0)}\right\rangle +\left|\Psi_{f}^{(1)}\right\rangle $
is given by

\begin{equation}
E1_{PNC}=\frac{\left\langle \Psi_{f}\right|\vec{D}\left|\Psi_{i}\right\rangle }{\sqrt{\left\langle \Psi_{f}\right.\left|\Psi_{f}\right\rangle \left\langle \Psi_{i}\right.\left|\Psi_{i}\right\rangle }}\,.\label{e1pnc}\end{equation}

The preliminary results we have obtained using this approach are given
in table \ref{e1pnc-table}. These calculations have been carried
out in the Dirac-Coulomb approximation with an universal Gaussian
basis consisting of $13s$, $12p$, $11d$ and $7f$ function for
$Cs$ and $13p$, $12p$, $11d$ and $8f$ functions for $Ba^{+}$.
Our results for $Cs$ is in reasonable agreement with linear relativistic
CCSD(T) calculation \cite{blundell-e1pnc} which yields $E1_{PNC}=0.0908(9)\times10^{-11}$
in the same units as our calculation.

\begin{table}

\caption{\label{e1pnc-table}$E1_{PNC}$ matrix elements for $Cs$ and $Ba^{+}$}

\begin{center}\begin{tabular}{ccc}
\hline 
Atom&
Transition&
$E1_{PNC}$($iea_{0}(Q_{W}/-N)$)\tabularnewline
\hline
\hline 
$Cs$&
$6s\,^{2}S_{1/2}\longrightarrow7s\,^{2}S_{1/2}$&
$0.910\times10^{-11}$\tabularnewline
&
&
\tabularnewline
$Ba^{+}$&
$6s\,^{2}S_{1/2}\longrightarrow5d\,^{4}D_{3/2}$&
$2.05\times10^{-11}$\tabularnewline
\hline
\hline 
&
&
\tabularnewline
\end{tabular}\end{center}
\end{table}

\section{Conclusion}

It is clear that considerable progress has been made during the past
three decades on the relativistic many-body theory of atoms. However
there are some open problems in this field. Perhaps the two areas
that deserve most attention is the future are :

(i) Relativistic multi-reference theories to treat a wide variety
of open shell heavy atoms including rare-earths. Work in this area
is in its infancy \cite{ishikawa-2004}.

(ii) Incorporation of QED effects in a systematic way in the framework
of relativistic many-body theory.

One can indeed look forward to exciting new developments in relativistic
electronic structure of atoms in the coming decade.

\begin{verse}
\textbf{Acknowledgements} : Some of the computations presented in
this review were performed on CDAC's Teraflop Supercomputer Param
Padma in Bangalore and in our group's Xeon PC cluster which was procured
from the BRNS project \# 2002/37/12/BRNS. Previous members of our
group, in particular, Dr. Angom Dilip Singh, Dr. Holger Merlitz, Dr.
Uttam Sinha Mahapatra, Dr. Geetha Gopakumar and Dr. Sonjoy Majumder
have made important contributions to our effort on relativistic many-body
theory of atoms. 
\end{verse}

\end{document}